\documentclass[review,number,sort&compress]{elsarticle}
\usepackage{subcaption}
\usepackage{graphicx}
\usepackage{amssymb}
\usepackage{color}
\usepackage{url}
\usepackage{ulem}
\usepackage[figuresright]{rotating}

\journal{Applied Radiation and Isotopes}

\begin{document}

\begin{frontmatter}

%% use the tnoteref command within \title for footnotes;
%% use the tnotetext command for theassociated footnote;
%% use the fnref command within \author or \address for footnotes;
%% use the fntext command for theassociated footnote;
%% use the corref command within \author for corresponding author footnotes;
%% use the cortext command for theassociated footnote;
%% use the ead command for the email address,
%% and the form \ead[url] for the home page:
%% \title{Title\tnoteref{label1}}
%% \tnotetext[label1]{}
%% \author{Name\corref{cor1}\fnref{label2}}
%% \ead{email address}
%% \ead[url]{home page}
%% \fntext[label2]{}
%% \cortext[cor1]{}
%% \address{Address\fnref{label3}}
%% \fntext[label3]{}

%\title{A source-based fast-neutron facility for precision irradiations}}
\title{Tagging fast neutrons from an $^{241}$Am/$^{9}$Be source}

%% use optional labels to link authors explicitly to addresses:
%% \author[label1,label2]{}
%% \address[label1]{}
%% \address[label2]{}

\author[lund,ess]{J.~Scherzinger}
\author[glasgow]{J.R.M.~Annand}
\author[arktis]{G.~Davatz}
\author[lund,ess]{K.G.~Fissum\corref{cor1}}
\ead{kevin.fissum@nuclear.lu.se}
\author[arktis]{U.~Gendotti}
\author[ess,midswe]{R.~Hall-Wilton}
\author[m4]{A.~Rosborg}
\author[lund]{E.~H\aa kansson}
\author[arktis]{R.~Jebali\fnref{fn2}}
\author[ess]{K.~Kanaki}
\author[m4]{M.~Lundin}
\author[ess,m4]{B.~Nilsson}
\author[m4,sweflo]{H.~Svensson}

\address[lund]{Division of Nuclear Physics, Lund University, SE-221 00 Lund, Sweden}
\address[ess]{Detector Group, European Spallation Source ESS AB, SE-221 00 Lund, Sweden}
\address[glasgow]{University of Glasgow, Glasgow G12 8QQ, Scotland, UK}
\address[arktis]{Arktis Radiation Detectors Limited, 8045 Z\"{u}rich, Switzerland}
\address[m4]{MAX IV Laboratory, Lund University, SE-221 00 Lund, Sweden}
\address[midswe]{Mid-Sweden University, SE-851 70 Sundsvall, Sweden}
\address[sweflo]{Sweflo Engineering, SE-275 63 Blentarp, Sweden}

\cortext[cor1]{Corresponding author. Telephone:  +46 46 222 9677; Fax:  +46 46 222 4709}
\fntext[fn2]{present address: University of Glasgow, Glasgow G12 8QQ, Scotland, UK}

\begin{abstract}
Shielding, coincidence, and time-of-flight measurement techniques are employed
to tag fast neutrons emitted from an $^{241}$Am/$^{9}$Be source resulting in
a continuous polychromatic energy-tagged beam of neutrons with energies up to
7~MeV. The measured energy structure of the beam agrees qualitatively with 
both previous measurements and theoretical calculations.
\end{abstract}


\end{frontmatter}

\section{Introduction}
\label{section:introduction}

\noindent
Fast neutrons are important probes of matter and diagnostic tools 
~\cite{walker82,homeland03,FNDA06,FNDA11,ESS09,chandra10,lyons11,chandra12,
peerani12, FNASS13,islam13,lewis13,tomanin14,lewis14}. 
Sources of fast neutrons for 
controlled irradiations include nuclear reactors, particle accelerators, and 
radioactive sources. Drawbacks associated with nuclear reactors and particle 
accelerators include their accessibility and availability, as well as the very 
high cost per neutron. In contrast, radioactive sources provide neutrons with 
a substantially lower cost per neutron. Drawbacks associated with radioactive 
sources include the complex mixed field of radioactive decay products which 
complicate the experimental situation. As a first step towards developing a 
source-based fast-neutron irradiation facility, we have employed well-understood 
shielding, coincidence, and time-of-flight (TOF) measurement techniques to attenuate 
and subsequently unfold the mixed decay-product radiation field provided by an 
$^{241}$Am/$^{9}$Be (hereafter referred to as Am/Be) source, resulting in a 
polychromatic energy-tagged neutron beam.

\section{Apparatus}
\label{section:apparatus}

\subsection{Am/Be source}
\label{subsection:ambe_source}

\begin{figure} %figure 01
\begin{center}
\resizebox{0.50\textwidth}{!}{\includegraphics{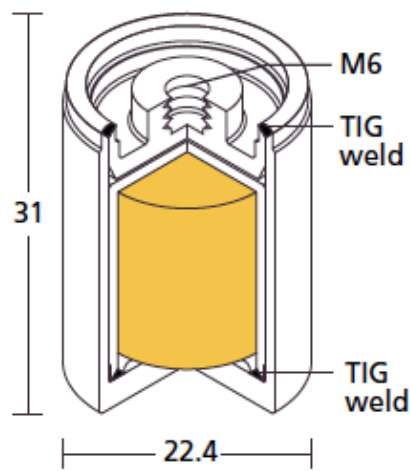}}
\caption{\label{figure:ambe_source}
The Am/Be source (figure from Ref.~\cite{hightechsoltd}). Dimensions in mm. 
The yellow shaded volume at the central core of the capsule corresponds to the 
Am/Be.  (For interpretation of the references to color in this figure caption, 
the reader is referred to the web version of this article.)
}
\end{center}
\end{figure}

The heart of the irradiation facility consists of a (nominal) 18.5~GBq Am/Be 
radioactive source~\cite{hightechsoltd}. This source is a mixture of americium 
oxide and beryllium metal contained in an X.3 capsule\footnote{
An X.3 capsule is a tig-welded, double-layered, stainless-steel cylinder 
approximately 30~mm (height) $\times$ 22~mm (diameter).}
(see Fig.~\ref{figure:ambe_source}). 

Radioactive $^{241}$Am has a half-life of 432.2 years and decays 
via $\alpha$ emission (5 different energies averaging $\sim$5.5~MeV) to 
$^{237}$Np. The dominant energy of the resulting background gamma-rays from 
the decay of the intermediate excited states in $^{237}$Np is $\sim$60~keV. 
$^{237}$Np has a half-life of over 2 million years. $^{9}$Be is stable. 

Fast neutrons are produced when the decay $\alpha$ particles interact with 
$^{9}$Be. Depending on the interaction and its kinematics, $^{12}$C and a free 
neutron may be produced. The resulting free-neutron distribution has a maximum 
value of about 11~MeV and a sub-structure of peaks whose energies and relative 
intensities vary depending upon the properties of the Am/Be source containment
capsule and the size of the $^{241}$AmO$_2$ and Be particles in the powders 
employed -- see the detailed discussion presented in Ref.~\cite{lorch73}. In 
general, approximately $\sim$25\% of the neutrons emitted have an energy of 
less than $\sim$1~MeV with a mean energy of $\sim$400~keV~\cite{hightechsoltd}. 
The average fast-neutron energy is $\sim$4.5~MeV. Both the gamma-ray and neutron 
dose rates at a distance of 1~m from our unshielded source in the X.3 capsule 
were measured to be 11~$\mu$Sv/h, for a total unshielded dose rate of 
22~$\mu$Sv/h. The unshielded source has been independently determinated to 
emit (1.106 $\pm$ 0.015)~$\times$~10$^6$ neutrons per second nearly 
isotropically~\cite{natphyslab}.

The kinematics and the reaction cross section for the $^{9}$Be($\alpha,n$)
interaction determine the state of the recoiling $^{12}$C nucleus produced in 
the reaction. The calculations of Vijaya and Kumar~\cite{vijaya73} (for 
example) suggest that the relative populations of the ground/first/second 
excited states for the recoiling $^{12}$C nucleus are 
$\sim$35\%/$\sim$55\%/$\sim$15\%. If the recoiling $^{12}$C nucleus is left in 
its first excited state, it will promptly decay to the ground state via the 
isotropic emission of a 4.44~MeV gamma-ray. Mowlavi and 
Koohi-Fayegh~\cite{mowlavi04} as well as Liu~{\it et al.}~\cite{liu07} have 
measured $R$, the 4.44~MeV $\gamma$-ray to neutron ratio for Am/Be, to be 
approximately 0.58.  Again, this is seemingly dependent upon the Am/Be capsule 
in question. Regardless, almost 60\% of the neutrons emitted by an Am/Be source
are accompanied by a prompt, time-correlated 4.44~MeV $\gamma$-ray.  We exploit 
this property of the source to determine neutron TOF and thus kinetic 
energy by measuring the elapsed time between the detection of the 4.44~MeV 
$\gamma$-rays and the detection of the fast neutrons. Note that by employing 
this technique, we necessarily restrict our available ``tagged" neutron energies 
to a maximum value of $\sim$7 MeV as 4.44~MeV of the reaction $Q$-value are 
``lost" to the de-excitation gamma-ray.

\subsection{YAP:Ce 4.44~MeV gamma-ray trigger detectors}
\label{subsection:yaps}

\begin{figure} %figure 02
\begin{center}
\rotatebox{90}{
\resizebox{0.50\textwidth}{!}{\includegraphics{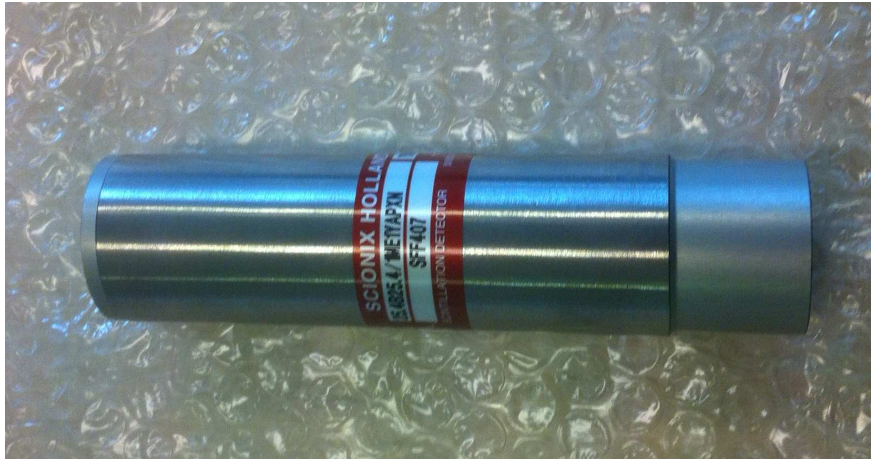}}
}
\caption{\label{figure:yap_detector}
Photograph of a YAP detector. A 1" (diameter) $\times$ 1" (height) crystal
has been mounted on a 1" (diameter) $\times$ 10~cm (length) PMT.
}
\end{center}
\end{figure}

The 2 YAP:Ce\footnote{
YAP:Ce stands for Yttrium Aluminum Perovskit:Cerium (YAlO$_{3}$, Ce$^{+}$ 
doped).
} 
fast ($\sim$5~ns risetime) gamma-ray trigger detectors (hereafter referred to 
as YAPs) were provided by Scionix~\cite{scionix}. A detector (see 
Fig.~\ref{figure:yap_detector}) consisted of a cylindrical 1" (diameter) $\times$ 
1" (height) YAP crystal~\cite{moszynski98} coupled to a 1" Hamamatsu Type R1924 
photomultiplier tube (PMT)~\cite{hamamatsu} operated at about $-$800~V. Gains 
for the YAP detectors were set using a YAP event trigger and standard gamma-ray 
sources. Typical energy resolution obtained for the 662~keV peak of $^{137}$Cs 
using such a detector was about 10\%.  YAP:Ce is radiation hard and quite 
insensitive to neutrons of all energies, which makes it ideal for detecting 
gamma-rays within the large fast-neutron field of the Am/Be source. We stress 
that because of their small volume, the YAP detectors were not used for 
spectroscopy, but simply to trigger on any portion of the energy deposited by 
the 4.44~MeV gamma-rays emitted by the source. A 3~mm thick Pb sleeve placed 
around the source (see Sec.~\ref{subsection:configuration}) to attenuate the 
high intensity 60~keV gamma-ray field and a 350 keV$_{ee}$ discriminator 
threshold proved to be an effective combination for the YAP detection of these 
4.44~MeV gamma-rays.

\subsection{NE-213 fast-neutron and gamma-ray liquid-scintillator detector}
\label{subsection:ne213_detector}

\begin{figure} %figure 03
\begin{center}
\resizebox{1.00\textwidth}{!}{\includegraphics{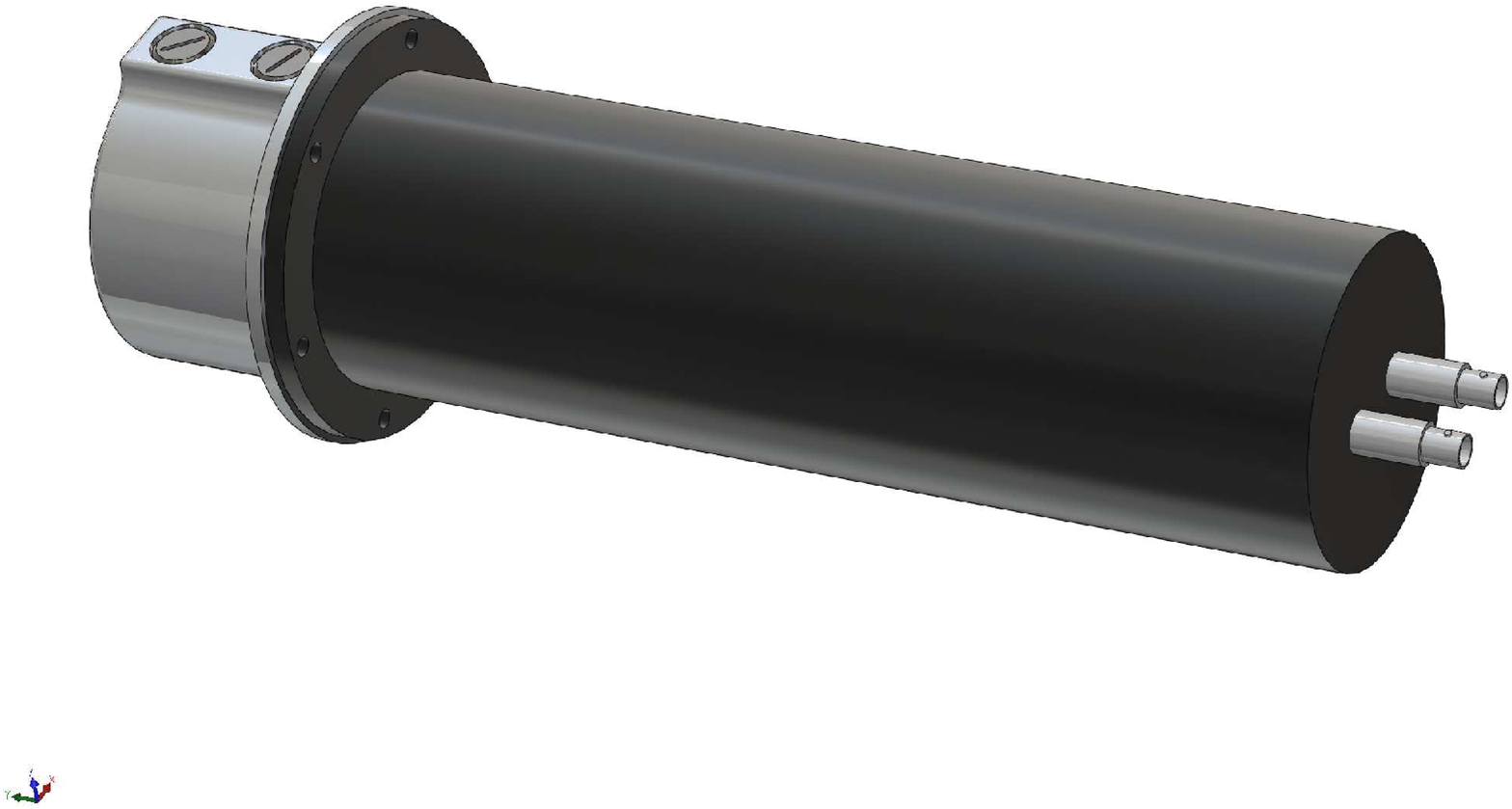}}
\resizebox{0.75\textwidth}{!}{\includegraphics{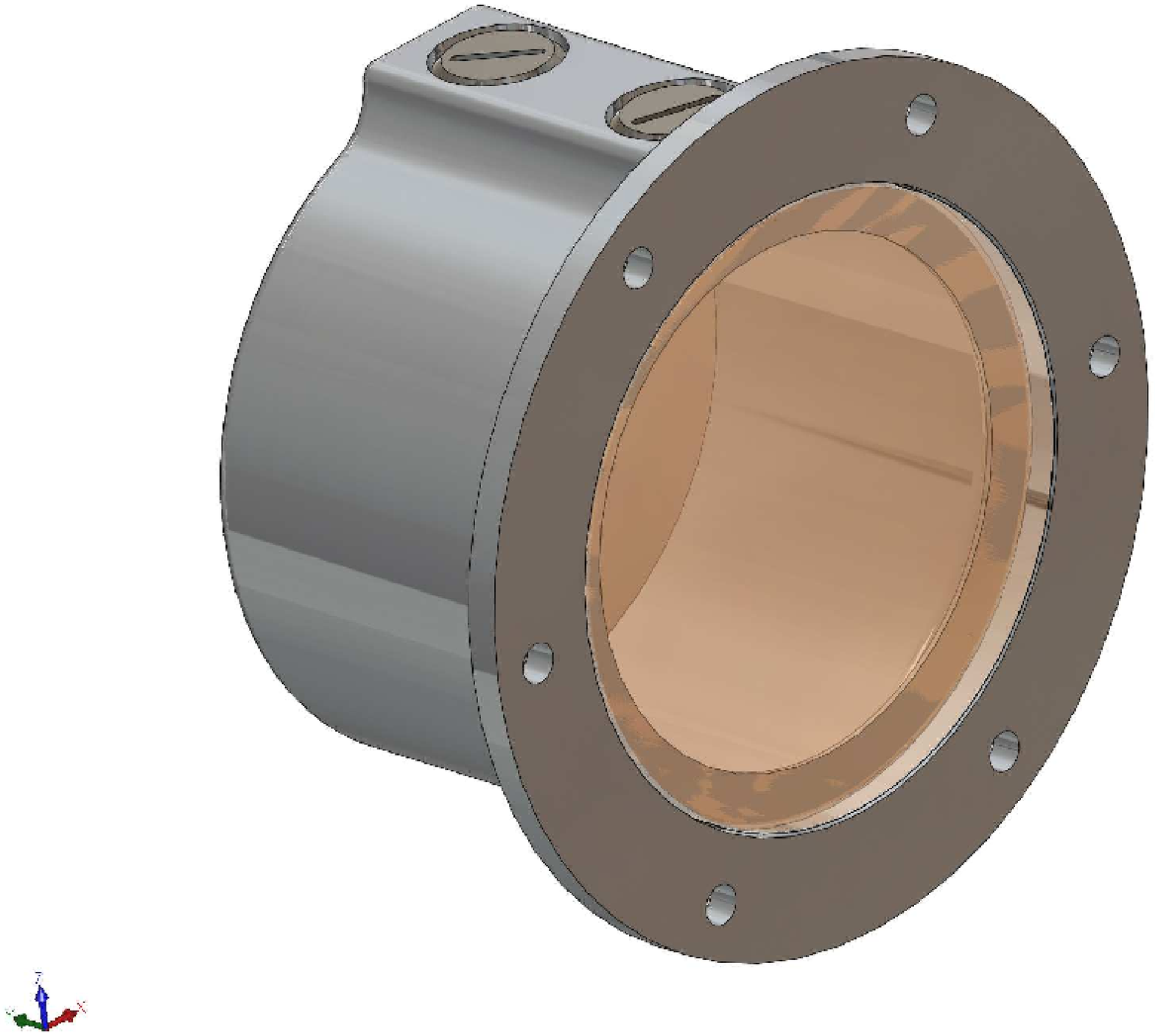}}
\caption{\label{figure:ne213_detector}
CAD drawing of the NE-213 detector. Top panel: the detector. The large black 
cylinder to the right is the magnetically shielded 3" ET Enterprises 9821KB 
photomultiplier-tube assembly. The small gray cylinder to the left contains 
the NE-213. Bottom panel: an enlargement of the small gray cylinder ``cup". 
The screws on top allow for the filling or draining of the liquid cylinder. 
A borosilicate-glass window (light brown) serves as the optical boundary. 
(For interpretation of the references to color in this figure caption, the 
reader is referred to the web version of this article.)
}
\end{center}
\end{figure}

The NE-213~\cite{ne213} fast-neutron and gamma-ray detector employed in this
work is shown in Fig.~\ref{figure:ne213_detector}. A 3~mm thick cylindrical 
aluminum cell with a depth of 62~mm and a diameter of 94~mm housed the NE-213.
The inside of the cell was treated with xylene-solvent withstanding 
EJ-520~\cite{ej520} titanium dioxide reflective paint. The cell was sealed 
with a 5~mm thick borosilicate glass plate~\cite{borosilicate} attached using 
Araldite 2000$+$~\cite{araldite} glue, which is highly resistant to both 
temperature and chemicals. The penetrations into the cell were closed with 
M-8 threaded aluminum plugs with 20~mm diameter heads and sealed with 14~mm 
diameter Viton O-rings~\cite{viton}. The assembled cell was filled with the 
nitrogen-flushed NE-213 using a nitrogen gas-transfer system. 

After the cell was filled, the borosilicate glass window was coupled to a 
cylindrical PMMA UVT lightguide~\cite{pmma} with a height of 57~mm and a 
diameter of 72.5~mm. The lightguide wall was painted with water-soluble 
EJ-510~\cite{ej510} reflective paint. The lightguide was then pressure-coupled 
to a spring-loaded, magnetically shielded 3~inch ET Enterprises 9821KB PMT 
assembly~\cite{et9821kb} operated at about $-$2000 V. In order to ensure the 
reproducibility of the behavior of the detector over an extended period of 
time rather than maximize light transmission, optical grease was not used in 
the assembly.  Gain for the NE-213 detector was set using an NE-213 detector 
event trigger and a set of standard gamma-ray sources together with the 
prescription of Knox and Miller~\cite{knox72}.

\subsection{Configuration}
\label{subsection:configuration}

\begin{figure} %figure 04
\begin{center}
\resizebox{1.00\textwidth}{!}{\includegraphics{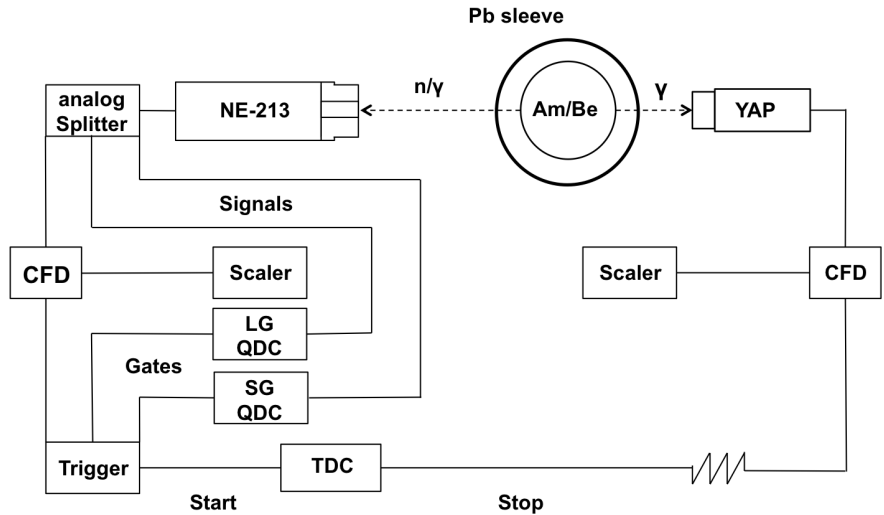}}
\caption{\label{figure:SetUp}
A simplified overview of the experimental setup (not to scale). The Am/Be source, 
the Pb sleeve, a single YAP detector, and a NE-213 detector are all shown together 
with a block electronics diagram.
}
\end{center}
\end{figure}

A block diagram of the experiment configuration is shown in 
Fig.~\ref{figure:SetUp}. The Am/Be source was placed so that its 
cylindrical-symmetry axis corresponded to the vertical direction in the lab at 
the center of a 3~mm thick cylindrical Pb sleeve (with the same orientation) 
to attenuate the 60~keV gamma-rays associated with the decay of 
$^{241}$Am\footnote{
The half-value layer for Pb for 60~keV gamma-rays is $<$1~mm.}.
A YAP detector was placed with its crystal approximately 5~cm from the Am/Be 
source at source height. The crystal orientation was such that its cylindrical
symmetry axis also corresponded to the vertical direction in the lab. This 
%detector was free-running, and triggered overwhelmingly on the 4.44~MeV 
detector triggered overwhelmingly on the 4.44~MeV 
gamma-rays radiating from the source which came from the decay of the first 
excited state of $^{12}$C.  A NE-213 detector was placed approximately 68~cm 
from the Am/Be source at source height. The cylindrical symmetry axis of the 
NE-213 detector pointed directly at the center of the source. This detector 
%was also free-running, and 
triggered on both 4.44~MeV gamma-rays and fast neutrons 
coming from the source, as well as cosmic rays and room background\footnote{
Room background consisted primarily of 2.23~MeV gamma-rays associated with
neutron capture on the hydrogen in the water and paraffin 
used as general radiation shielding about the source.}.

\subsection{Electronics and data acquisition}
\label{subsection:electronics_and_daq}

The analog signals from the YAP trigger detector and the NE-213 detector were 
passed to LRS 2249A and 2249W CAMAC charge-to-digital converters (QDCs) and
PS 715 NIM constant-fraction (timing) discriminators. The resulting logic 
signals from the discriminators were passed to LRS 2228A CAMAC time-to-digital 
converters (TDCs) and LRS 4434 scalers. These signals were recorded on an 
event-by-event basis for offline processing using a LINUX PC-based data-acquisition 
(DAQ) system exploiting the ROOT~\cite{root} data-analysis framework. Connections 
to VME and CAMAC crates were respectively facilitated by a SBS 616 PCI-VME bus 
adapter and a CES 8210 CAMAC branch driver. In YAP calibration mode, signals from 
a YAP detector were periodically employed to trigger the DAQ and thus monitor the 
gains of the YAP detectors. In TOF mode, signals from the NE-213 detector were 
used to trigger the DAQ so that the gain of the NE-213 detector was continuously 
monitored. The NE-213 detector QDCs included a 60~ns short-gated (SG) QDC and a 
500~ns long-gated (LG) QDC, both of which opened 25~ns before the analog pulse 
arrived. The NE-213 detector also provided the start trigger for the TOF TDC. The 
YAP trigger provided the stop trigger for the TOF TDC. By triggering our 
data-acquisition system on the NE-213 detector, we avoided unnecessary deadtime 
processing events seen only by the YAPs.  Two particular source-related occurrences 
were of special interest: 1) a fast neutron detected in the NE-213 detector 
starting the TOF TDC with the corresponding 4.44~MeV gamma-ray detected in 
the YAP detector stopping it; and 2) prompt, time-correlated gamma-ray pairs 
emitted from the source being detected in coincidence in the NE-213 and YAP 
detectors (see below). 

\section{Results}
\label{section:results}

\begin{figure} %figure 05
\begin{center}
\resizebox{1.0\textwidth}{!}{\includegraphics{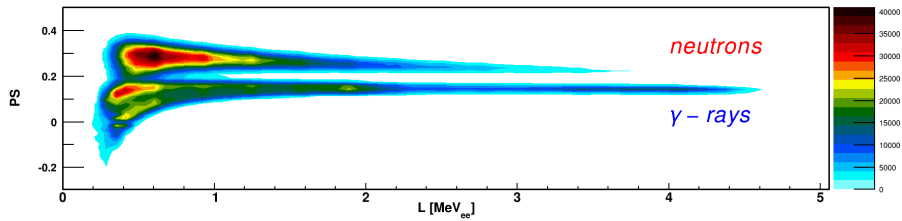}}
\caption{\label{figure:PSD}
A contour plot of pulse shape (PS) versus total energy deposited in the LG QDC 
(L) for correlated fast-neutron/gamma-ray events in the NE-213 detector. The 
upper distribution corresponds to neutrons while the lower band corresponds to 
gamma-rays.
}
\end{center}
\end{figure}

Figure~\ref{figure:PSD} shows a contour plot of the energy deposited in the 
NE-213 detector as a function of ``pulse shape" (PS, see below) versus ``L" 
(the energy deposited in the LG QDC). PS was calculated using the 
``tail-to-total" method~\cite{jhingan08,lavagno10,pawelczak13}; namely, the 
difference in the energies registered by the LG and SG QDCs was normalized to 
the energy registered by the LG QDC. As the NE-213 scintillator responded 
differently\footnote{
In the liquid scintillator NE-213, gamma-ray scintillations are fast while 
neutron-associated scintillations have pronounced slow components. Analysis of 
the time structure of the scintillation components leads to particle 
identification (PID) and is known as pulse-shape discrimination (PSD).}
to gamma-ray and fast-neutron events, the two distinct distributions appeared
in the PS versus L contour plot. Particle identification (PID) based solely 
upon the pulse-shape discrimination (PSD) characteristics of the NE-213 
detector was good, although some overlap between the distributions existed in 
the vicinty of PS~$\sim$~0.2.

\begin{figure} %figure 06
\resizebox{1.00\textwidth}{!}{\includegraphics{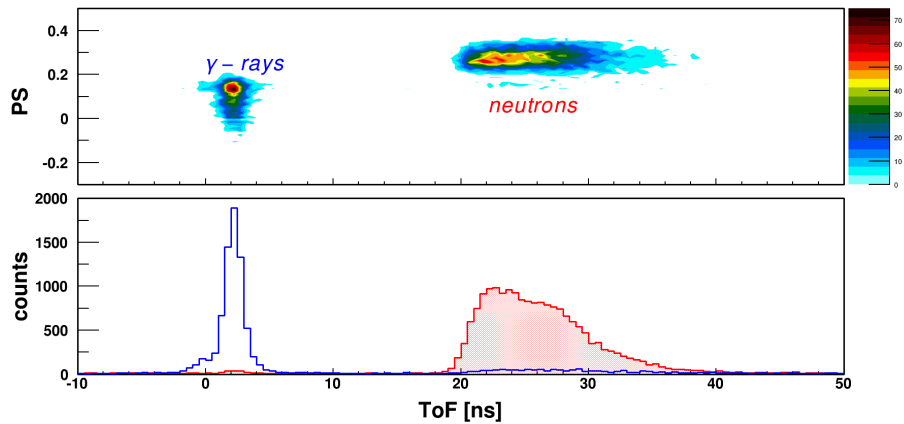}}
\caption{\label{figure:TOF}
Time-of-flight distributions. 
The top panel shows a contour plot of PS versus time-of-flight (ToF).  The 
bottom panel shows the projection of events from the top panel onto the 
time-of-flight axis. Events identified as gamma-rays in Fig.~\ref{figure:PSD} 
(the $\gamma$-flash) are presented in the sharp (blue) unshaded histogram, 
while events identified as neutrons are presented in the (red) shaded 
histogram. (For interpretation of the references to color in this figure 
caption, the reader is referred to the web version of this article.)
}
\end{figure}

Figure~\ref{figure:TOF} presents the time-of-flight distribution of the data 
shown in Fig.~\ref{figure:PSD}. 
No software cut on L was applied in mapping the data from Fig.~\ref{figure:PSD} 
to Fig.~\ref{figure:TOF}. The hardware threshold was 250 keV$_{ee}$. The top 
panel shows a contour plot of PS versus time-of-flight. Time-of-flight based 
PID is clearly excellent. The bottom panel shows the projection of events from 
the top panel onto the time-of-flight axis subject to a PS $=$ 0.19 cut to 
separate neutrons from gamma-rays. The sharp (blue) unshaded peak centered at 
about 2~ns is known as the ``$\gamma$-flash"\footnote{
The instant of the production in the source of the correlated pair of events
which produce the time-of-flight data is known as ``T$_{0}$" and is located 
at a time-of-flight of 0~ns.}.  
The gamma-flash corresponds to a pair of prompt, time-correlated gamma-rays 
produced in the source which triggered both the NE-213 detector and the YAP 
detector. The $\sim$1.8~ns FWHM of the gamma-flash is consistent with the timing 
jitter on our PMT signals.  The tail of events to the right of the gamma-flash 
corresponds to non-prompt gamma-rays\footnote{
A non-prompt gamma-ray can result from inelastic neutron scattering.} 
and randoms (see below).  The broad (red) shaded peak centered at about 25~ns 
corresponds to time-correlated 4.44~MeV gamma-ray/fast-neutron pairs where the 
fast neutron triggered the NE-213 detector while the 4.44~MeV gamma-ray 
triggered the YAP detector. A neutron with time-of-flight measured in this 
manner has been tagged. The very low level of background consists of randoms. 
Random events arose when the NE-213 detector started the time-of-flight 
measurement, but no correlated stop was received from the YAP. Typical random 
events included cosmic rays, room background, Am/Be neutrons not correlated 
with a 4.44~MeV gamma-ray, and Am/Be neutrons where the 4.44~MeV gamma-ray was 
missed due to YAP inefficiency or geometry.

\begin{figure} %figure 07
\begin{center}
\resizebox{1.00\textwidth}{!}{\includegraphics{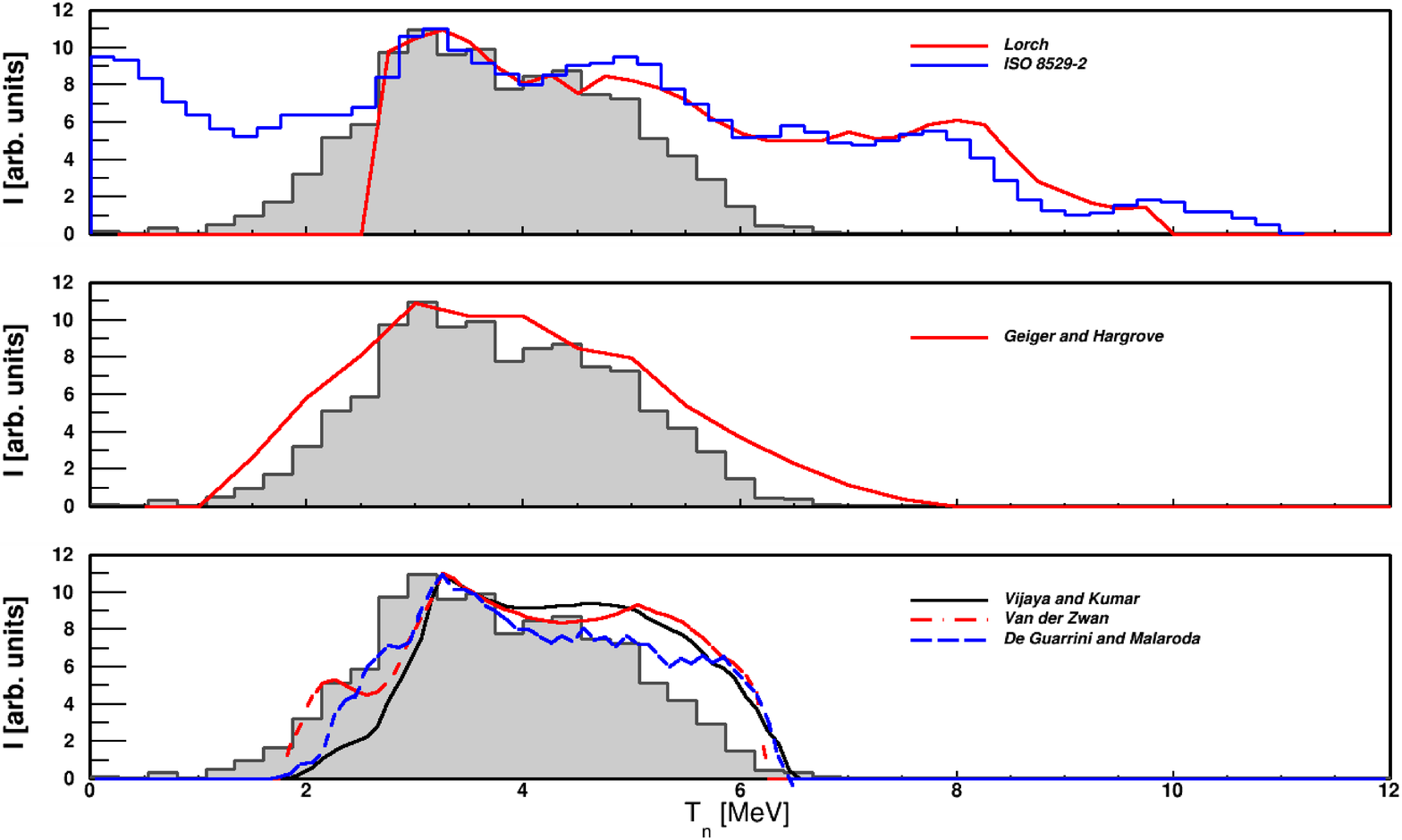}}\\
\caption{\label{figure:E_kin}
Our results (the gray shaded histogram) and comparisons. Top panel: to the 
full-energy neutron spectrum (red line) of Lorch~\cite{lorch73} and the 
ISO~8529-2 reference neutron radiation spectrum for Am/Be (blue open histogram). 
Middle panel: to tagged-neutron results obtained by Geiger and 
Hargrove~\cite{geiger64} (red line). Bottom panel: to theoretical calculations 
of the tagged-neutron spectrum. The solid black line represents the calculation 
of Vijaya and Kumar~\cite{vijaya73}, the red dashed line represents the 
calculation of Van~der~Zwan~\cite{vanderzwan68}, and the blue dot-dashed 
line represents the calculation of De~Guarrini and Malaroda~\cite{deguarrini71}. 
(For interpretation of the references to color in this figure caption, the 
reader is referred to the web version of this article.)
}
\end{center}
\end{figure}

Figure~\ref{figure:E_kin} shows our tagged-neutron results together with 
previous results, the ISO~8529-2 reference neutron radiation spectrum for 
Am/Be\footnote{
While we employ the reference spectrum in our discussion of results, the
interested reader may prefer Refs.~\cite{marsh95} and \cite{chen14}.
}, and theoretical calculations. Our data represent yield -- they have 
not been corrected for neutron-detection efficiency or detector acceptance. 
In all 3 panels, the maximum values of the spectra at $\sim$3~MeV have been 
normalized to our distribution. The reference neutron radiation spectrum
is shown in the top panel together with the full-energy neutron spectrum of 
Lorch~\cite{lorch73} which is widely quoted in conjunction with work with 
Am/Be sources. Agreement between the Lorch data and the reference spectrum 
is very good between 2.5~MeV and 10~MeV. The reference spectrum shows some 
strength above 10~MeV which Lorch did not observe. Our data show no strength 
above $\sim$7~MeV due to the neutron-tagging procedure -- 4.44~MeV 
potentially available to the neutron are ``lost" to the creation of the 
de-excitation gamma-ray. This is neither an acceptance nor an efficiency 
effect, it is purely energetics. The reference spectrum shows considerable 
strength below 2.5~MeV. Our data also show some strength in this region.
The Lorch data do not. The sharp cutoff at about 2.5~MeV in the Lorch 
data is not directly discussed in the reference, but based upon its 
appearance in spectra from several different sources all measured with the 
same apparatus, we attribute it to an analysis threshold cut as it lies well 
above their quoted neutron-detector threshold of 1~MeV. Our hardware 
threshold was 250~keV$_{ee}$ corresponding to a neutron energy of 
$\sim$1.3~MeV, and no analysis threshold cut was employed. The agreement 
between our data, those of Lorch, and the reference spectrum between 2.5 
and 5~MeV (in the region of overlap) is excellent. The method of tagging 
the 4.44~MeV de-excitation gamma-ray and a comparable Am/Be source\footnote{
Their source capsule was slightly smaller and emitted about 50\% more neutrons
per second.} 
were employed by Geiger and Hargrove~\cite{geiger64} in obtaining the 
results shown in the middle panel. Both the neutrons and the gamma-rays from 
their source were detected in Naton 136 plastic scintillators. Agreement with 
our results is very good.  We attribute the small difference in the strengths 
observed in the two measurements to neutron-detection efficiency and 
acceptance effects which we do not consider. We attribute the relative 
broadening of their measured neutron distribution with respect to ours to 
their quoted poorer than 12\% energy resolution for neutron detection, 
which based on the numbers quoted in their manuscript, we gather was 
calculated at 2~MeV. At 2~MeV, based upon our gamma-flash FWHM of 1.8~ns,
time-of-flight path length of 0.675~m, and detector half-depth of 3.1~cm,
our energy resolution was 11\%. At 4~MeV, our energy resolution was 19\%.  
The three independent theoretical calculations of the 
tagged-neutron yield shown in the bottom panel come from Vijaya and 
Kumar~\cite{vijaya73}, Van~der~Zwan~\cite{vanderzwan68}, and De~Guarrini and 
Malaroda~\cite{deguarrini71}. The details of these calculations are beyond the 
scope of this paper, but clearly all three are in reasonable agreement both 
with each other as well as our results. We conclude we are tagging 
neutrons.

\section{Summary}
\label{section:summary}

We have employed shielding, coincidence, and time-of-flight measurement techniques 
to tag fast neutrons emitted from an Am/Be source as a first step towards developing 
a source-based fast-neutron irradiation facility. The resulting continuous
polychromatic energy-tagged neutron beam has a measured energy structure that
agrees qualitatively with both previous measurements and theoretical calculations. 
We conclude that our approach works as expected, and anticipate that it can 
provide a cost-effective means for detector characterization and tests of shielding.  
We note that this technique will work equally well for all Be-compound neutron 
sources.

\section*{Acknowledgements}
\label{acknowledgements}

We thank the Photonuclear Group at the MAX IV Laboratory for providing access 
to their experimental hall and Am/Be source. We acknowledge the support of the 
UK Science and Technology Facilities Council.

\bibliographystyle{elsarticle-num}

\end{document}